Karol Izydor Wysokiński
Instytut Fizyki
Uniwersytet Marii Curie – Skłodowskiej
Lublin

# Nadprzewodnictwo – pierwsze 100 lat

# Superconductivity – the first 100 years

Z okazji stulecia odkrycia zjawiska nadprzewodnictwa przypominam najważniejsze fakty z pierwszego okresu oraz próby teoretycznego jego zrozumienia. Okazuje się, że nad objaśnieniem zjawiska pracowały najtęższe umysły pierwszej połowy XX w. Udało się to dopiero w 1957 roku Bardeenowi, Cooperowi i Schriefferowi. Teoria BCS objaśniała wszystkie znane fakty i przewidywała nowe, co szybko zostało potwierdzone doświadczalnie, a teoria zaakceptowana jako poprawne objaśnienie zjawiska. Odkrycia ostatniego ćwierćwiecza stawiają nowe wyzwania dla teorii. Okazuje się, że prosty model BCS nie wystarcza do zrozumienia właściwości nowych nadprzewodników, które nazywamy niekonwencjonalnymi albo egzotycznymi. Chociaż nadprzewodniki znalazły wiele ważnych zastosowań w różnych dziedzinach życia, to wciąż nie spełniły olbrzymich nadziei w nich pokładanych. Mimo to, ta gałąź nauki rozwija się bardzo dynamicznie i wciąż fascynuje nowe generacje fizyków.

On the occasion of centenary of superconductivity discovery I remind some facts from the first period and attempt to understand the phenomenon. It turns out that most famous physicists of the first half of XX century have tried to solve the puzzle. Bardeen, Cooper and Schrieffer succeeded in 1957. BCS theory successfully described all known facts and offered new predictions, which soon have been confirmed experimentally contributing to the widespread acceptance of the theory. New discoveries of the last quarter of the century put new requirements for the theory. It turns out that simple BCS model is not enough to understand new unconventional or exotic superconductors. Even though the superconductors have found many important applications in various branches of technology they have not yet fulfilled the hopes. In spite of that the scientific studies of superconductors develop vividly and fascinate new generations of physicists.

## 1. Wstęp – dwie relacje z dnia narodzin

Odkrycie nadprzewodnictwa w 1911 roku poprzedziła wielka i bardzo dobrze udokumentowana [1] batalia o uzyskanie niskich temperatur i skroplenie He. Znamy znacznie mniej szczegółów związanych z odkryciem zjawiska nadprzewodnictwa. Co więcej, niektóre z relacji niezbyt dokładnie zgadzają się ze sobą [2, 3]. Jedna z nich [2] została, wg jej autora, zasłyszana od technika ściśle współpracującego z Heike Kamerlingh-Onnesem.

Szukając odpowiedzi na kluczowe na początku XX w. pytanie o zachowanie oporu elektrycznego czystych metali w najniższych temperaturach badano najpierw przewodnictwo platyny i złota, a potem rtęci, gdyż ten metal łatwo było uzyskać w bardzo czystym stanie. Zdawano sobie bowiem sprawę, że obecność domieszek zwiększa wartość oporu właściwego metali. Pomiary wykonywano w ten sposób, że ustalano wartość temperatury, a następnie mierzono opór próbki. Według relacji Gerrita Jana Flima opisanej przez Jacobusa de Nobela [2], gdy oziębiono próbkę rtęci do temperatury nieco poniżej 4.2 K, jej opór spadł do zera. Pomiary wykonywał Gilles Holst, który wynik ten zinterpretował jako zwarcie elektryczne w układzie pomiarowym. Cały układ rozebrano i pomiar powtórzono. I znowu zwarcie. Według relacji naocznego świadka, odkrycie nadprzewodnictwa zawdzięczamy przypadkowi. Gdy Holst po raz kolejny mierzył opór rtęci w temperaturze poniżej 4.2 K, uczeń Szkoły Konstruktorów Przyrządów, którego zadaniem było utrzymywanie stałego ciśnienia w kriostacie, zdrzemnął się. Ciśnienie par helu w kriostacie nieco wzrosło. Wzrosła też temperatura wrzenia helu przewyższając temperaturę przemiany nadprzewodzącej rtęci ($T_c$ = 4.14 K), co spowodowało pojawienie się skończonej wartości oporności próbki.

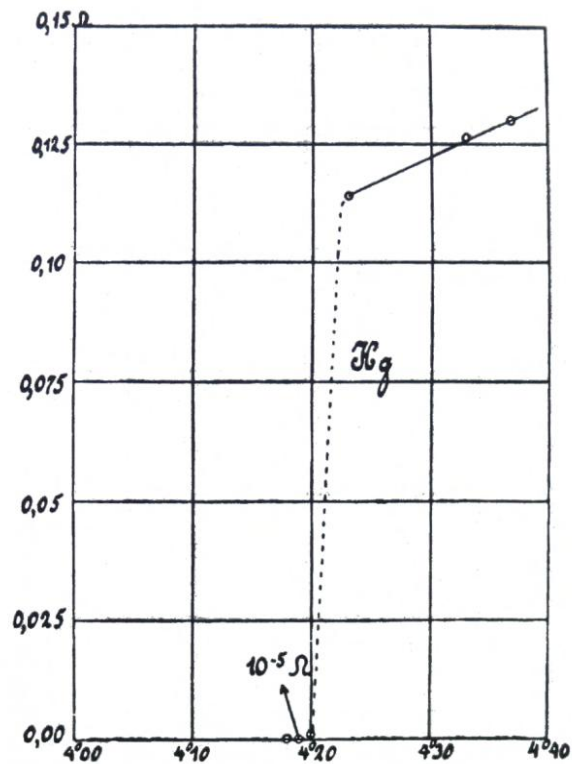

Rys. 1 Oryginalny wykres z pracy H. Kamerlingha-Onnesa [4] przedstawiający zależność oporności rtęci od temperatury.

Wydaje się jednak, że ta sympatyczna historyjka nie ma mocnego pokrycia w faktach. Tak przynajmniej twierdzą autorzy innej relacji [3], którzy dokładnie przestudiowali notatki laboratoryjne H. Kamerlingha-Onnesa z okresu odkrycia nadprzewodnictwa (zawierające jednak szereg błędnych dat). Dokładne odczytywanie mało czytelnych zapisów dziennika laboratoryjnego z 1911 roku pozwala, wg autorów relacji znaleźć wpis z dnia 8 kwietnia, stwierdzający „[oporność] rtęci bliska zeru". W często ostatnio cytowanym oryginale „kwik nagenoeg nul". Według tej relacji w Leidzie zdawano sobie sprawę z możliwości uzyskania takiego wyniku, choć nie z jego konsekwencji. Tego dnia (tzn. 8 kwietnia 1911 r.) pomiary rozpoczęto o 7 rano, a Kamerlingh-Onnes przybył do laboratorium o 11.20. Także według tej relacji pomiary wykonywał Gilles Holst. Zmierzył on opór rtęci i złota w temperaturze 4.3 K (czyli tuż powyżej $T_c$ rtęci). Potem obniżono ciśnienie pary i uzyskano temperaturę około 3 K. Dokładnie o godzinie 4.00 po południu ponownie zmierzono opór rtęci i złota. Tym razem w T = 3 K i to wtedy okazało się, że oporność rtęci wynosi prawie zero.

Dokładne odczytanie zapisków Kamerlingha-Onnesa z końca tego dnia przynosi kolejną sensacyjną obserwację, której waga naukowa jednak nie dotarła do świadomości badaczy. W dzienniku Kamerlingha-Onnesa tak zostało to zapisane [3]: *„Just before the lowest temperature [about 1.8 K] was reached, the boiling suddenly stopped and was replaced by evaporation in which liquid visibly shrank. So, a remarkably strong evaporation*

*at the surface"*[1]. Zapis ten nie pozostawia wątpliwości, że zaobserwowano nadciekłą przemianę helu, pod normalnym ciśnieniem zachodzącą w temperaturze ok. 2.2 K.

Pomiary oporności rtęci i innych pierwiastków, były wielokrotnie powtarzane w następnych tygodniach i miesiącach. Znany, historyczny już wykres zależności oporności rtęci (rys. 1) od temperatury został wykonany 26 października 1911 roku i opublikowany przez Kamerlingha-Onnesa bez uwzględnienia rzeczywistych wykonawców pomiarów [4]. W tym czasie Kamerlingh-Onnes oraz współpracownicy zaczęli zdawać sobie sprawę z wagi odkrycia i potencjalnych możliwości zastosowań zjawiska. W kontekście moich osobistych zainteresowań silnym nieporządkiem w nadprzewodnikach, chciałbym odnotować fakt, że już 20 czerwca 1911 roku Kamerlingh-Onnes po dyskusji z Holstem zdecydował się przeprowadzić pomiary oporności rtęci domieszkowanej złotem i kadmem. Ponieważ opór tak przygotowanych próbek w niskich temperaturach także nagle malał do zera, zaniechano dalszych systematycznych badań. W laboratorium Kamerlingha-Onnesa odkryto kilka innych nadprzewodników w tym ołów, który zmieniał stan w 6 K oraz cynę o $T_c = 4$ K. Dopiero te odkrycia przekonały Kamerlingha-Onnesa i jego współpracowników, że mają do czynienia z nowym zjawiskiem, a nie jakimś trudnym do wykrycia artefaktem aparaturowym.

Heike Kamerlingh-Onnes otrzymał w 1913 r. Nagrodę Nobla z fizyki za skroplenie helu. Laudatio Komitetu Noblowskiego brzmiało *„for his investigations on the properties of matter at low temperatures which led,* inter alia*, to the production of liquid helium"*[2].

Przez kolejnych dwadzieścia lat znakiem rozpoznawczym nadprzewodnictwa nowych materiałów był zanik oporności do niemierzalnie niskich wartości poniżej temperatury przemiany $T_c$.

## 2. Powody, dla których Kamerlingh-Onnes badał oporność metali

W literaturze przedmiotu wymienia się przynajmniej dwa ważne powody, dla których Kamerlingh-Onnes badał oporność metali w coraz niższych temperaturach. Pierwszy z nich to chęć posiadania termometrów do stosowania w zakresie bardzo niskich temperatur. Zwykle używano termometrów gazowych. Jednak liniowa zależność oporności wielu metali od temperatury obserwowana w okolicy temperatury wrzenia ciekłego wodoru zachęcała do konstrukcji niezależnego miernika temperatury.

Drugi powód, to chęć sprawdzenia hipotez Kelvina na temat oporności metali w najniższych temperaturach. Argumentacja Kelvina była dość ciekawa [5]. Twierdził on, że jeżeli potrzeba bardzo wysokich temperatur, aby szkło stało się przewodzące, oznacza to, że w temperaturach pokojowych nośniki prądu, elektrony, są w szkle zamrożone. W metalach, które dobrze przewodzą w temperaturach pokojowych elektrony mogą – podobnie jak w szkle – ulec zamrożeniu (przy atomach) w bardzo niskich temperaturach. Wtedy przewodnictwo elektryczne znikałoby.

Inny możliwy scenariusz przewidywał, że w niskich temperaturach drgania oscylatorów (fonony) powodujące rozpraszanie elektronów ulegają zamrożeniu, więc elektrony nie rozpraszają się, co oznacza, że oporność znika, a przewodnictwo jest nieskończone. Wiedziano już wtedy, że obecność domieszek powoduje wzrost oporności elektrycznej. Aby badać najczystsze metale, wybrano rtęć, którą można było bardzo dobrze oczyścić metodą kolejnych destylacji i umieścić w kapilarnej rurce do badań niskotemperaturowych [5]. Wynik Kamerlingha-Onnesa zdawał się, zatem potwierdzać tę drugą ewentualność i w pierwszej pracy o odkryciu zjawiska tak zostało to zinterpretowane, mimo iż spadek oporności był skokowy.

---

1 *Tuż przed uzyskaniem najniższej temperatury [około 1.8 K], wrzenie gwałtownie ustało i zostało zastąpione przez parowanie, w trakcie czego ciecz widocznie skurczyła się (zmniejszyła objętość). Co za silne parowanie z powierzchni.*
2 *Za badania właściwości materii w niskich temperaturach, które doprowadziły m.in. do uzyskania ciekłego helu.*

## 3. Zrozumieć zjawisko – wielcy fizycy XX w. w konfrontacji z zagadką stanu nadprzewodzącego

Wielu znakomitych fizyków epoki usiłowało zrozumieć zjawisko zerowania się oporu elektrycznego. Wszystkie próby były bezowocne. Felix Bloch, którego osiągnięcia w związku z kwantowym opisem ruchu elektronów w periodycznym potencjale kryształów są dobrze znane studentom fizyki, zajmował się także zjawiskiem nadprzewodnictwa. Chyba frustracja związana z brakiem postępów spowodowała, że powiedział [6]: *„the only theorem about superconductivity that can be proved is that any theory of superconductivity is refutable"*[3]. Stwierdzenie to nazywano czasami żartobliwie „twierdzeniem Blocha".

Sytuacja miała się zmienić po odkryciu w 1933 przez W. Meissnera i R. Ochsenfelda [7], że nadprzewodnik oziębiony poniżej temperatury $T_c$ zawsze wypycha ze swego wnętrza pole magnetyczne. Oznacza to, że we wnętrzu nadprzewodnika $\vec{B} = 0$. Proste zastosowanie teorii Drude przewodnictwa metali i równań elektrodynamiki (równań Maxwella) pozwala uzyskać zależność [8-9].

$$\nabla^2 \frac{d\vec{B}}{dt} = \frac{1}{\lambda^2} \frac{d\vec{B}}{dt}, \qquad (1)$$

która pokazuje, że zmienne pole magnetyczne jest „tłumione" na odległości $\lambda$ w materiale, którego opór właściwy znika. We wzorze (1) $\lambda^2 = m_e/n_s e^2$, a $m_e$ oznacza masę elektronu, $e$ – jego ładunek, natomiast $n_s$ gęstość kondensatu. Genialne odkrycie braci Fritza i Heinza Londonów w 1934 roku polegało na obserwacji, że równanie takie w odniesieniu do pola $\vec{B}$, a nie jego czasowych zmian $\frac{d\vec{B}}{dt}$ poprawnie opisuje wyniki doświadczenia Meissnera-Ochsenfelda.

Fritz London był, w zgodnej opinii wielu badaczy, tym który jako pierwszy uważał zjawisko nadprzewodnictwa za przejaw kwantowej koherencji makroskopowego układu. Co więcej argumentował, że może to być nadprzewodnik o wymiarach centymetrów czy kilometrów. Jako pierwszy nazwał je makroskopowym zjawiskiem kwantowym. Znakomity opis osiągnięć braci Londonów można znaleźć w artykule napisanym z okazji stulecia nadprzewodnictwa zatytułowanym „Zapomniani bracia" i opublikowanym w Physics World [6]. Autor cytuje J. Bardeena, który stwierdził *„it was Fritz London who first recognised that superconductivity and superfluidity result from manifestations of quantum phenomena on the scale of large objects"*[4].

Chyba wszyscy najwięksi fizycy XX w. w taki lub inny sposób zajmowali się objaśnieniem zjawiska nadprzewodnictwa. Pierwszy wielki sukces to niewątpliwie wspomniana teoria Londonów. Na kolejny należało poczekać piętnaście lat. Była to termodynamiczna teoria Landaua przemian fazowych ciągłych sformułowana w latach trzydziestych i zastosowana przez Ginzburga i Landaua w 1950 r. do opisu nadprzewodnictwa [10]. Teoria Ginzburga-Landaua, podobnie jak i teoria Londonów jest teorią fenomenologiczną. W odróżnieniu od poprzedniej uwzględnia energię powierzchniową związaną z granicą pomiędzy obszarem normalnym i nadprzewodzącym. Kondensat jest tu opisywany za pomocą zespolonego, zależnego od punktu w przestrzeni parametru porządku $\Psi(r) = |\Psi(r)|e^{i\theta(r)}$. Jedno z równań teorii, jakie uzyskujemy z minimalizacji energii swobodnej nadprzewodnika w zewnętrznym polu magnetycznym opisywanym, analogicznie jak w mechanice kwantowej za pomocą potencjału wektorowego $\vec{A}(\vec{r})$ ma postać podobną do nieliniowego równania Schrödingera

$$\frac{1}{2m^*}\left(-i\hbar\nabla + e^*\vec{A}\right)^2\Psi(r) + a\Psi(r) + b|\Psi(r)|^2\Psi(r) = 0. \qquad (2)$$

---

[3] *Jedynym twierdzeniem na temat nadprzewodnictwa, które można udowodnić jest to, że dowolną teorię nadprzewodnictwa można obalić.*
[4] *To Fritz London jako pierwszy zauważył, że nadprzewodnictwo i nadciekłość są przejawem kwantowych zjawisk w skali makroskopowej.*

Drugie równanie przypomina równanie na gęstość prądu prawdopodobieństwa

$$\vec{j}(\vec{r}) = i\frac{e^*h}{4\pi m^*}(\psi^*\nabla\psi - \psi\nabla\psi^*) - \frac{(e^*)^2}{m^*}|\psi(\vec{r})|^2\vec{A}(\vec{r}), \qquad (3)$$

ale jest równaniem na gęstość prądu elektrycznego w nadprzewodniku. Prąd $\vec{j}(\vec{r})$ poprzez równanie Maxwella określa wartość pola magnetycznego w nadprzewodniku. Warunki brzegowe, jakie nałożone są na „makroskopową" funkcję falową $\Psi(\vec{r})$ zapewniają, że znika składowa prądu nadprzewodzącego normalna do powierzchni próbki, natomiast składowa pola magnetycznego styczna do powierzchni nadprzewodnika jest ciągła na granicy nadprzewodnik – ośrodek normalny. Równanie (3) teorii Ginzburga-Landaua przechodzi w równanie Londonów, jeśli założyć, że faza funkcji falowej jest stała, a jej moduł utożsamić z gęstością kondensatu, $n_s$.

Według relacji Ginzburga [11], występujące w tej teorii parametry $m^*$ i $e^*$ o wymiarach masy i ładunku były powodem poważnych dyskusji pomiędzy Ginzburgiem i Landauem. Problem polegał na tym, że w teorii kwantowej z niezmienniczością na cechowanie wiąże się zasada zachowania ładunku, a w powyższych równaniach $e^*$ mogło przyjmować dowolną wartość, nawet zależną od temperatury lub położenia. Landau utrzymywał, że $e^* = e$, natomiast Ginzburg szacował, że wartość ta może być większa niż ładunek elektronu $e^* = (2 - 3)e$.

Przełom związany z mikroskopową teorią nadprzewodnictwa nastąpił w połowie lat pięćdziesiątych. W 1956 roku Leon Cooper pokazał, że jeśli pomiędzy dwoma elektronami w metalu o energiach wyższych od energii Fermiego $E_F$ występuje nawet infinitezymalnie słabe oddziaływanie przyciągające, to tworzą one stan związany. Oznacza to, że całkowita energia tych elektronów jest niższa niż $2E_F$ – czyli minimalnej energii, jaką mogą one posiadać w metalu. Wynik ten wskazujący na niestabilność powierzchni Fermiego względem oddziaływań przyciągających pomiędzy elektronami utorował drogę pierwszej poprawnej mikroskopowej teorii nadprzewodnictwa. Została ona opublikowana w 1957 roku przez J. Bardeena, L. Coopera i R. Schrieffera [12] i jest znana jako teoria BCS. Autorzy zaproponowali postać wariacyjnej funkcji falowej układu $N$ elektronów w metalu

$$|\Psi_{\text{BCS}}\rangle = \prod_{\vec{k}}\left(u_{\vec{k}} + v_{\vec{k}}c^+_{k\uparrow}c^+_{-k\downarrow}\right)|0\rangle \qquad (4)$$

Wielkości $u_{\vec{k}}$, $v_{\vec{k}}$ są, w ogólności zespolonymi parametrami wariacyjnymi, $|0\rangle$ oznacza stan próżni kwantowej, a operatory $c^+_{k\uparrow}\left(c^+_{-k\downarrow}\right)$ są operatorami kreacji elektronu stanie o wektorze falowym $\vec{k}$ i spinie $\uparrow$ ($-\vec{k}, \downarrow$). Iloczyn przebiega po wszystkich dozwolonych stanach kwantowych elektronów. Drugi wyraz w ostatnim wzorze opisuje tworzenie się par elektronowych o zerowym pędzie środka masy, które nazywamy parami Coopera.

Teoria BCS przewiduje, zgodnie z szeregiem wyników doświadczalnych, pojawienie się w widmie elektronów metalu, szczeliny energetycznej w otoczeniu poziomu Fermiego o wartości

$$\Delta \approx \sim 2\hbar\omega_D e^{-\frac{1}{N(0)V}}. \quad (5)$$

Wynik powyższy uzyskuje się przy założeniu, że efektywne oddziaływanie przyciągające pomiędzy elektronami ma w bliskim otoczeniu powierzchni Fermiego stałą wartość $V$. $N(0)$ jest gęstością stanów metalu normalnego na poziomie Fermiego, natomiast $\omega_D$ jest częstością Debye'a fononów. Częstość Debye'a pojawia się, gdyż w teorii tej zakłada się, że to oddziaływanie elektronów z fononami prowadzi do efektywnego przyciągania pomiędzy fermionami.

Teoria poprawnie przewidywała zależność ciepła właściwego od temperatury i jego skok w $T = T_c$ i szereg innych dobrze znanych właściwości nadprzewodników. Z tego powodu została bardzo dobrze przyjęta. Wkrótce pojawiły się dodatkowe doświadczenia jednoznacznie potwierdzające jej słuszność. Jednym z ważniejszych była obserwacja tzw. wierzchołka koherencyjnego w widmie jądrowego rezonansu magnetycznego (NMR) [13].

Opisując pierwsze 100 lat nadprzewodnictwa można dowolną liczbę stron tekstu przeznaczyć na omawianie różnych osiągnięć teoretycznych i doświadczalnych i nie uzyskać oczekiwanego wyniku. Próba podsumowania osiągnięć tej dziedziny wiedzy została podjęta pod koniec lat sześćdziesiątych. Dwutomowe dzieło o objętości prawie 1400 stronic znakomicie opisuje pewien rozdział badań nadprzewodnictwa, które nazywamy klasycznym [14].

Wydaje się, że z okazji stulecia odkrycia zjawiska warto, choćby pobieżnie, omówić (nieudane) próby zrozumienia nadprzewodnictwa przez innych wielkich fizyków tamtych czasów. Pokaże nam to skalę trudności, jaką udało się pokonać formułując poprawny opis zjawiska. W pracach historycznych często wymienia się, obok wspomnianego już Blocha, takich uczonych jak: Einstein, Bohr, Brillouin, Born czy Feynman. Nie wszyscy z nich wiele pisali na temat nadprzewodnictwa, ale wszyscy mniej lub bardziej intensywnie starali się je zrozumieć.

Znana jest tylko jedna praca Einsteina z 1922 r., która od paru lat dostępna jest w angielskim tłumaczeniu w postaci preprintu zamieszczonego w arXiv [15]. Pracę tę napisał Einstein z okazji czterdziestej rocznicy profesury H. Kamerlingha-Onnesa w Lejdzie. Einstein wprowadza w niej koncepcję molekularnych łańcuchów i twierdzi, że prąd nadprzewodzący związany jest z cyklicznym ruchem elektronów w zamkniętych łańcuchach. Jego propozycja doświadczenia sprawdzającego przewidywania teorii przyczyniła się do szybkiej jej falsyfikacji. Einstein przewidywał, że prąd nadprzewodzący nie będzie płynął poprzez złącze zbudowane z dwu różnych nadprzewodników, gdyż nie posiadają one wspólnych łańcuchów molekularnych. Doświadczenie Kamerlingha-Onnesa wykonane zanim praca Einsteina została wydrukowana pokazało, że przez złącze ołowiu i cyny płynie prąd nadprzewodzący, gdyż nie zauważono mierzalnej oporności takiego złącza. Znakomite omówienie wczesnej historii nadprzewodnictwa i roli Einsteina w określeniu pewnych kierunków badań w Lejdzie można znaleźć w pracy T. Sauera [16].

Wydaje się, że także W. Pauli, który nie opublikował żadnej pracy na temat nadprzewodnictwa był dość mocno zaangażowany w badania nadprzewodnictwa. Wynika to z jego listu do Bohra, w którym czytamy „*w sprawie nadprzewodnictwa, nie doszedłem do żadnych konkretnych wyników*" [17]. Natomiast znamy wyobrażenia Bohra na temat zjawiska. Uważał on, że nadprzewodnictwo związane jest ze skoordynowanym ruchem całej sieci elektronowej. Jeśli wszystkie elektrony w metalu tworzą sieć, to rozpraszanie jednego z nich na domieszkach czy jonach nie jest możliwe. Przejście do stanu normalnego w podejściu Bohra mogło być związane z „topnieniem" tej sieci elektronowej [17].

R. Feynman napisał jedną pracę na temat nadprzewodnictwa [18] i to przeglądową. Jednak w wywiadzie w 1988 stwierdził „*Centralnym problemem dla mnie było nadprzewodnictwo i spędziłem tak wiele czasu, aby je zrozumieć. [...] Nie opublikowałem nigdy niczego na ten temat i w moich publikacjach jest znaczna przerwa spowodowana właśnie próbami rozwiązania problemu nadprzewodnictwa, czego mi się nie udało dokonać*" (cytowanie za [19]). Feynman analizował problem z wykorzystaniem poprawnego Hamiltonianu w ramach teorii zaburzeń i wywnioskował, że jeżeli istnieje rozwiązanie problemu nadprzewodnictwa, to z pewnością nie jest ono możliwe do uzyskania w ramach teorii zaburzeń. I miał rację, gdyż poprawna teoria nadprzewodnictwa nie jest teorią zaburzeniową, o czym świadczy nieanalityczny (typu exp(-*1/x*) tzn. nierozwijalny w szereg potęgowy względem *x*) charakter zależności parametru szczeliny energetycznej Δ od potencjału oddziaływania *V* w równaniu (5).

Interesująca analiza historii nadprzewodnictwa oraz zainteresowania zjawiskiem R. Feynmana znajduje się w pracy [20]. Autorzy uzasadniają, częściowo bazując na osobistych kontaktach, że Feynman do końca życia żywo interesował się zjawiskiem nadprzewodnictwa, choć w jednym z wywiadów przyznawał: „*W moim umyśle pojawiła się blokada emocjonalna dotycząca tego zagadnienia, kiedy więc dowiedziałem się o pracy BCS, to przez długi czas nie mogłem się zmusić, żeby ją przeczytać*"[19]. Po odkryciu nadprzewodników wysokotemperaturowych w 1986 roku, Feynman przewidywał, że materiałem o najwyższej temperaturze przejścia będzie nadprzewodnik na bazie skandu (Sc). Wydaje się, że nie miał racji, chyba że przyroda ma dla nas kolejne niespodzianki w tym zakresie.

Warto wspomnieć, że pierwsza próba Landaua fenomenologicznego opisu nadprzewodnictwa z 1933 roku też okazała się błędna. Landau rozważał możliwość pojawienia się w stanie równowagi termodynamicznej spontanicznych prądów $\vec{j}$. W tej teorii postulował, aby przedstawić energię swobodną nadprzewodnika *F*, jako sumę energii swobodnej normalnego metalu, $F_N$, co w tej teorii oznaczało materiał z zerowym prądem oraz wyrazów zawierających kolejne parzyste (aby energia była skalarem niezależnym od kierunku prądu) potęgi $\vec{j}$

$$F(\vec{j}) = F_N + \frac{a}{2}(\vec{j})^2 + \frac{b}{4}(\vec{j})^4 + \dots . \qquad (6)$$

Prąd równowagowy $\langle \vec{j} \rangle$ obliczamy minimalizując energię swobodną. Uzyskana zależność prądu $|\vec{j}|$ od temperatury typu $(T_c - T)^{1/2}$ nie była zgodna z istniejącymi wynikami doświadczalnymi i teoria została odrzucona.

## 4. Kluczowe doświadczenia

W rozwoju teorii nadprzewodnictwa kluczową rolę odegrały 2 doświadczenia. Pierwsze to wspomniane wypychanie pola magnetycznego. Ponieważ zjawisko było odwracalne, pozwoliło to wywnioskować, że nadprzewodnictwo jest stanem równowagowym układu i zastosować równania termodynamiki takich stanów. Obliczając różnicę entropii układu w stanie normalnym (*N*) i nadprzewodzącym (*S*) znajdujemy

$$S_N - S_S = -\mu_0 H_c \left( \frac{dH_c}{dT} \right)_H \qquad (7)$$

co wskazuje, że przemiana jest ciągła, gdy zachodzi w zerowym zewnętrznym polu magnetycznym (gdy *H* = 0, to $H_c$ = 0, przy *T* = $T_c$) natomiast jest nieciągła, gdy pole *H* > 0, gdyż wtedy przejście stan normalny – nadprzewodzący zachodzi w temperaturze nieco niższej niż $T_c$, a wtedy $H_c \neq 0$. Co więcej – z monotonicznego zmniejszania się pola krytycznego $H_c$ ze wzrostem temperatury, aż do wartości zerowej w $T = T_c$ (oznacza to, że prawa strona wzoru (7) jest dodatnia) wnioskujemy, że faza normalna jest fazą o wyższej entropii. Faza nadprzewodząca jest więc fazą o wyższym stopniu uporządkowania.

Drugie z ważnych doświadczeń na drodze do poprawnej mikroskopowej teorii to pomiar temperatury krytycznej dwu próbek tego samego materiału wykonanych z różnych jego izotopów. Odkryte zjawisko izotopowe [21], wykazujące zależność temperatury przemiany od masy $M$ izotopu w postaci

$$T_C \sim M^{-\alpha} \qquad (8)$$

z wartością $\alpha \approx 0,5$ było mocnym argumentem za ważną rolą układu jonów i oddziaływania elektronów z jonami. W tym samym czasie teoretycznie stwierdzono, że w wyniku oddziaływania elektronów z drganiami sieci może pojawić się pomiędzy nimi efektywne oddziaływanie przyciągające [22].

## 5. Nowe odkrycia – nowe wyzwania

Jeśli mówimy o nowych odkryciach i wyzwaniach w fizyce ciała stałego, a szczególnie w kontekście nadprzewodnictwa to tak się składa, że zwykle są to odkrycia doświadczalne, które stanowią wyzwania dla teorii. Tylko w kilku przypadkach nowe zachowania zostały najpierw przewidziane teoretycznie, a później potwierdzone doświadczalnie. Wspomnijmy tu o dwu bardzo ważnych i brzemiennych w skutki wynikach teoretycznych.

Pierwszy, to analiza teorii Ginzburga-Landaua przeprowadzona w 1957 roku przez Abrikosova [23] w granicy $\lambda > \xi$ ($\lambda$ - głębokość wnikania, $\xi$ - długość koherencji). Istnienie ujemnej energii powierzchniowej w tej granicy wskazywało na stabilność stanu nadprzewodzącego powyżej dolnego pola krytycznego $H_{c_1}$ aż do wartości $H_{c_2}$. W ten sposób przewidziano istnienie nadprzewodników II rodzaju o dużych wartościach górnego pola krytycznego. Wkrótce okazało się, że domieszkowanie nadprzewodników I rodzaju zmienia charakter zależności pomiędzy $\lambda$ i $\xi$ z $\lambda \ll \xi$ na $\lambda \leq \xi$ lub $\lambda > \xi$ i typ nadprzewodnika z I na II rodzaj. Wynika to stąd, że długość koherencji $\xi_0$ w nadprzewodnikach z niewielką drogą swobodną $l$ nośników przyjmuje wartość $\xi \approx \sqrt{\xi_0 l} < \xi_0$.

Drugim ważnym odkryciem teoretycznym jest zjawisko tunelowania par Coopera przez cienkie bariery tunelowe powodujące przepływ prądu pomiędzy dwoma nadprzewodnikami. Doświadczenie tego typu było wykonane już przez Kamerlingha-Onnesa w kontekście teorii Einsteina, jednak wynik Josephsona [24] jest niezależny. Dobry opis okoliczności, w jakich zjawisko Josephsona zostało odkryte teoretycznie znajduje się we wspomnieniach P. W. Andersona [25]. Znane teraz pod nazwą „stałoprądowe zjawisko Josephsona" polega na przepływie prądu stałego o natężeniu

$$I_s = I_c \sin(\Delta \varphi) \qquad (9)$$

pomiędzy dwoma nadprzewodnikami, których funkcje falowe Ginzburga-Landaua charakteryzują się różnicą faz $\Delta \varphi$. Wynik był tak nieoczekiwany, że nawet J. Bardeen – twórca teorii BCS argumentował, że jest błędny, jako że w obszarze bariery znika amplituda par Coopera. Wkrótce okazało się, że to młody student miał rację, a zjawisko jest wykorzystywane w urządzeniach zwanych SQUID-ami (*od Superconducting QUantum Interference Device*) służącymi obecnie m.in. do bardzo precyzyjnych ($\sim 10^{-15}$ T) pomiarów pól magnetycznych.

Josephson wykazał także, że jeśli do nadprzewodników po obu stronach bariery tunelowej przyłożyć stałą różnicę potencjałów, to spowoduje ona zmianę w czasie różnicy faz pomiędzy nadprzewodnikami

$$\frac{d}{dt}\Delta\varphi = \frac{2e}{\hbar}V \qquad (10)$$

a co za tym idzie – przepływ przez złącze prądu zmiennego

$$I_s = I_c \sin\left(\Delta\varphi_0 + \frac{2e}{\hbar}Vt\right) \qquad (11)$$

o częstotliwości $\omega=2eV/\hbar$ i amplitudzie $I_c$. Parametr $I_c$ w obu wzorach oznacza krytyczną wartość prądu. To „zmiennoprądowe zjawisko Josephsona" jest wykorzystywane w metrologii, jako precyzyjny wzorzec napięcia.

Odkrycie zjawiska nadprzewodnictwa o tak niezwykłych właściwościach pobudziło wyobraźnię wielu badaczy. Już Kamerlingh-Onnes marzył o licznych zastosowaniach zjawiska. Potrzebne do tego są materiały charakteryzujące się możliwie wysokimi temperaturami przejścia $T_c$, wysokimi wartościami pól krytycznych $H_{c_2}$ i prądów krytycznych $I_c$. Poszukiwano więc wciąż nowych nadprzewodników, a każde odkrycie przynosiło nową nadzieję i nowe wyzwania opisu właściwości materiału.

W 1941 roku odkryto nadprzewodnictwo w związku NbN z $T_c$ = 16 K, a w 1953 r. $V_3Si$ o rekordowej wtedy temperaturze przemiany 17,5 K. Pierwszy nadprzewodzący drut został wykonany ze stopu NbTi na początku lat sześćdziesiątych. Przez wiele lat nadprzewodnikiem o rekordowo wysokiej temperaturze przejścia pozostawał $Nb_3Ge$ ($T_c \approx 23.4$ K) o strukturze krystalograficznej A15.

## 6. Ważne odkrycia nowych nadprzewodników

Odkrycia nowych nadprzewodników lub całych ich rodzin następują dość regularnie. Nie wszystkie one wzbudzają tak szerokie zainteresowanie jak nadprzewodzące tlenki miedzi znane pod nazwą „nadprzewodniki wysokotemperaturowe" [26]. Duże zainteresowanie wzbudziły odkrycia ostatnich kilkunastu lat: nadprzewodnictwa w $MgB_2$ ($T_c$ = 39 K), $Sr_2RuO_4$ ($T_c$ = 1,5 K) i domieszkowanych fullerydkach ($T_c$ = 43 K). O szeregu tych nadprzewodników można znaleźć informację w popularnych opracowaniach: na temat początkowych etapów badań nadprzewodników wysokotemperaturowych [27, 28], fullerydków [29], $MgB_2$ oraz $Sr_2RuO_4$ [30].

W tym miejscu należy przypomnieć dwa znaczące odkrycia polskich uczonych. W 1972 roku prof. Tadeusz Skośkiewicz odkrył nadprzewodnictwo [31] w wodorku palladu $PdH_x$ o temperaturze krytycznej około 12 K oraz odwrotne zjawisko izotopowe. Odwrotne zjawisko izotopowe oznacza, że wzrost masy izotopowej prowadzi do wzrostu temperatury krytycznej, co oznacza ujemną wartość współczynnika α we wzorze (8). W tym przypadku zamiana wodoru H deuterem D zwiększa $T_c$ materiału o kilkanaście procent [32]. W latach siedemdziesiątych bardzo wiele grup badawczych zajmowało się tym nadprzewodnikiem usiłując wyjaśnić jego wysoką temperaturę przejścia oraz odwrotny efekt izotopowy. Prace [31] i [32] uzyskały po kilkaset cytowań.

Drugie zaskakujące odkrycie, które należy wspomnieć jest autorstwa prof. Andrzeja Kołodziejczyka i polegało na współistnieniu nadprzewodnictwa i ferromagnetyzmu w $Y_4Co_3$ (albo $Y_9Co_7$) [33]. To znakomite odkrycie nie znalazło jednak tak wielkiego oddźwięku w literaturze, na jaki zasługuje. Problemy technologiczne z uzyskaniem dobrej jakości próbek i złożona struktura elektronowa i magnetyczna to możliwe przyczyny takiego stanu rzeczy. W materiale tym uporządkowanie ferromagnetyczne występuje poniżej 6-9 K, natomiast nadprzewodnictwo pojawia się w temperaturze 3 K. Wydaje się, że w odróżnieniu od czteroskładnikowych związków metali przejściowych [34] i innych nadprzewodników [35] w $Y_9Co_7$ te same elektrony są odpowiedzialne za oba typy uporządkowań. Współistnienie takie zaobserwowano ostatnio w $UGe_2$ oraz URhGe [36].

Od 3 lat uwagę badaczy przykuwają odkryte 5 lat temu nadprzewodniki zawierające żelazo [37,38]. Związek LaOFeP [37] miał temperaturę krytyczną 3,2 K i nie wzbudził większego zainteresowania. Dopiero, gdy ta sama grupa doniosła w 2008 roku, że domieszkowanie fluorem zwiększa temperaturę krytyczną związku $LaFeAsO_{1-x}F_x$ do 26 K, wzrastającą do 43 K pod ciśnieniem, nastąpiło gwałtowne zainteresowanie tą rodziną nadprzewodników. Najwyższą temperaturę krytyczną (55 K) posiada materiał $SmFeAsO_{1-x}F_x$. Istnieje pięć klas nadprzewodników zawierających żelazo. Podstawowe ich właściwości są zebrane w tabeli 1. Nadprzewodniki żelazowe, zarówno pniktydki, jak i chalkogenidki mają strukturę warstwową. Uważa się, że za nadprzewodnictwo odpowiadają warstwy pniktydowo-żelazowe. Na diagramie fazowym w zmiennych $T_c$ – stopień domieszkowania, stan nadprzewodzący sąsiaduje ze stanem antyferromagnetycznego metalu. Stany elektronowe w pobliżu powierzchni Fermiego mają głównie charakter 3d od orbitali żelaza. Powierzchnia Fermiego składa się z kilku płatów. Zwykle rozważa się dwie powierzchnie elektronowe i dwie dziurowe. Szereg wyników wskazuje, że symetria parametru porządku jest typu *s*, przy czym znak na powierzchniach dziurowych jest przeciwny niż na elektronowych. Stan o takiej symetrii zwykle oznacza się symbolem $s_\pm$. Spore wartości współczynnika efektu izotopowego dla żelaza mogą świadczyć o ważnej roli oddziaływania elektron-fonon.

Tabela 1. Charakterystyczne parametry dla kilku przedstawicieli nadprzewodników żelazowych [39]. $T_c$ jest temperaturą przejścia w stan nadprzewodzący $\Delta_{min}$ i $\Delta_{max}$ oznaczają minimalne i maksymalne wartości szczeliny nadprzewodzącej, γ jest współczynnikiem elektronowego ciepła właściwego w relacji $c_e = \gamma T$, $\alpha_{Fe}$ współczynnikiem zjawiska izotopowego przy zamianie izotopów Fe. $H_{c_2}(c,ab)$ oznacza górne pole krytyczne (dla pola wzdłuż osi *c* i w płaszczyźnie *ab*), $\lambda_{ab}$ jest głębokością wnikania pola w płaszczyźnie *ab*.

| Oznaczenie | Materiał | $T_c$ [K] | $\Delta_{min}$ [meV] | $\Delta_{max}$ [meV] | γ [mJ/ molK$^2$] | $\alpha_{Fe}$ | $H_{c2}$ [T] Ab | $H_{c2}$ [T] c | $\lambda_{ab}$ [nm] |
|---|---|---|---|---|---|---|---|---|---|
| 1111 | LaFeAsO | 26 | | | 4.1 | | | | ~300 |
| | SmFeAsO | 55 | 4.2 | 6.5 | | 0.34 | | | ~210 |
| 122 | KFe2As2 | 3.3 | | | 69 | | 1.25 | 4.40 | |
| | $Ba_{0,6}K_{0,4}Fe_2As_2$ | 38 | 5.8 | 12.3 | 7.7 | -0.18 ÷0.36 | 56 | 57 | 200 |
| 111 | LiFeAs | 18 | 3.1 | 3.1 | | | | | 210 |
| 11 | $Fe_{1+x}Se$ | 8 | | | 1.3-5.4 | 0.81 | | | |
| | $FeTe_{0,6}Se_{0,5}$ | 15 | 0.51 | 2.61 | | | 47 | 44 | ~600 |

## 7. Dlaczego nadprzewodnictwo fascynuje

W codziennym życiu spotykamy się z tarciem, zawsze musimy uwzględniać opory ruchu. Wiele zaś praw fizycznych, o których uczymy lub uczyliśmy się w szkole dotyczy sytuacji idealnych bez oporów ruchu. Trudno jest wyobrazić sobie obiekt, który raz rozpędzony do pewnej prędkości będzie się z tą prędkością poruszał tak długo aż nie zadziałamy na niego inną siłą.

Gdy uczymy się o elektryczności, stwierdzamy, że każdy przewodnik charakteryzuje się oporem zależnym od jego geometrii (a więc: długości i pola przekroju poprzecznego), rodzaju materiału i temperatury. Tylko w zadaniach pojawia się często jakże nierealne polecenie: „zaniedbać opór przewodów doprowadzających prąd do urządzenia". Każdy myślący uczeń czuje, że to jest bardzo poważne założenie.

A oto szereg materiałów oziębionych do odpowiednich dla każdego z nich temperatur przestaje stawiać opór. Najczulsze przyrządy nie wykazują spadku potencjału wzdłuż przewodnika. Opór jest zerowy. Mamy do czynienia z idealnym przewodnikiem (z nadprzewodnikiem). Obcowanie z czymś idealnym jest nie lada gratką. To pobudza wyobraźnię. Obok takiego zjawiska nie można przejść obojętnie.

Wystarczy rozejrzeć się, aby pobudzić wyobraźnię w jeszcze większym stopniu. Przecież naokoło nas jest pełno przewodów z prądem. Niezależnie gdzie jesteśmy, czy w domu, na wycieczce, w biurze czy w laboratorium. Wszędzie mamy do czynienia z prądem płynącym w miedziowych lub innych przewodach. Wszystkie one stawiają opór. Czasami jest to korzystna sytuacja (grzejnik elektryczny, żelazko itp.), ale w większości przypadków opór jest niekorzystny. Tracimy bardzo wiele energii, która ucieka do atmosfery i jest bezpowrotnie tracona. Szacuje się, że ¼ wytwarzanej w elektrowniach energii jest tracona podczas jej przesyłania. To są olbrzymie koszty – dodajmy dla gospodarki i środowiska. Zastąpienie wszystkich zwykłych przewodów nadprzewodnikami oznaczałoby olbrzymie oszczędności. Jedyny problem polega na tym, że znane obecnie nadprzewodniki należy chłodzić do temperatury około -200$^o$C. To kosztuje. Rachunek ekonomiczny jest więc trochę bardziej skomplikowany. Przesyłowe linie energetyczne, urządzenia do diagnostyki medycznej eliminujące operacje chirurgiczne, pociągi poruszające się na poduszce magnetycznej, czy beztarciowe łożyska to tylko niektóre z pomysłów, zresztą już realizowane.

Liczba publikacji może być miarą fascynacji zjawiskiem w środowisku naukowym i nadziei na jego zastosowania. Po odkryciu nadprzewodników wysokotemperaturowych oczekiwania i nadzieje były tak duże, że powstawały nowe instytuty naukowe, specjalistyczne czasopisma i przedsiębiorstwa usiłujące wytwarzać i sprzedawać urządzenia nadprzewodnikowe.

Jest jeszcze jeden aspekt fascynacji zjawiskiem i nowymi nadprzewodnikami. Dla badaczy chyba najważniejszy. Związany jest on z chęcią zrozumienia właściwości tych materiałów. A są one fantastycznie skomplikowane. Przyroda bardzo zazdrośnie strzeże swoich tajemnic. Może dlatego, że wciąż dokonywane są coraz to nowe odkrycia, 100 lat po jego pierwszej obserwacji zjawisko to pokazuje coraz piękniejsze swoje strony. Najbardziej fascynujące i trudne są problemy związane z odpowiedzią na pytanie o symetrie parametru porządku i o mechanizmy odpowiedzialne za nie w różnych klasach materiałów.

Chyba największe kontrowersje budzą nadprzewodniki wysokotemperaturowe. Złośliwi twierdzą, że tyle jest mechanizmów nadprzewodnictwa wysokotemperaturowego ilu fizyków teoretyków zajmujących się tą tematyką. Sądzę, że taka sytuacja wynika ze złożoności zjawiska. Mamy tu prawdopodobnie do czynienia z szeregiem oddziaływań pozostających względem siebie w delikatnej równowadze.

## 8. Zastosowania

Kamerlingh-Onnes pojął znaczenie odkrytego przez siebie zjawiska dla wytwarzania silnych pól magnetycznych w nadprzewodzących solenoidach. Jednak już na początku 1914 roku stwierdził, że jego marzenie o wytwarzaniu bardzo silnych pól będzie trudne do zrealizowania, gdyż silne prądy niszczyły nadprzewodnictwo. W rzeczywistości to wytworzone pole magnetyczne niszczy stan nadprzewodzący.

O zastosowaniach nadprzewodników myślano od samego początku. Wiele z pomysłów zostało zrealizowanych. Rzeczywistość jest jednak taka, że są one wciąż poniżej oczekiwań. W wielu sytuacjach rachunek ekonomiczny jest najważniejszym argumentem. Niezależnie od tej pesymistycznej oceny łatwo jest wymienić szereg urządzeń, które sprawnie funkcjonują dzięki temu, że udało się wyprodukować nadprzewodzące przewody elektryczne, przez które płynie prąd o olbrzymim natężeniu. O zastosowaniach nadprzewodników można znaleźć wiele informacji i nie będę tego wątku dalej rozwijał. Zainteresowanego zagadnieniem czytelnika odsyłam np. do artykułu [40], jaki ukazał się w Physics World. Wydaje się, a przynajmniej Autor tak chce wierzyć, że nadprzewodniki nie powiedziały swego ostatniego słowa.

Ostatnie lata charakteryzują się gwałtownym rozwojem mikroelektroniki i związanym z nim rozwojem komputerów. Najnowsze urządzenia elektroniczne oparte na krzemie są bliskie osiągnięciu kresu możliwości. Przewidywane standardowe zastosowania nadprzewodników w mikroelektronice i technice komputerowej związane z budową klasycznych komputerów z wykorzystaniem nadprzewodzących, super-szybkich elementów są bardzo zaawansowane. Zaawansowane są też prace nad budową nadprzewodzącego komputera wykonującego $10^{15}$, czyli tysiąc bilionów operacji zmiennoprzecinkowych (np. dodawanie liczb rzeczywistych) na sekundę. Elektronicy rozważają także szereg innych bardziej przyziemnych zastosowań nadprzewodników takich jak superszybkie przełączniki, konwertery analogowo-cyfrowe, wydajne generatory itp. Niestety, wspomniany rachunek ekonomiczny nie czyni tych propozycji poważnymi konkurentami dla istniejących urządzeń półprzewodnikowych.

Ostatnio coraz więcej się mówi o wykorzystaniu nadprzewodników do budowy komputerów kwantowych. Jest to obiecujący kierunek badań i zapewne będzie jedną z szybciej rozwijających się gałęzi fizyki ciała stałego w najbliższych latach. Komputer kwantowy jest takim urządzeniem, które w jednym akcie obliczeniowym uzyska wynik dla wszystkich liczb z danego przedziału, podczas gdy klasyczny komputer w takim samym akcie obliczeniowym może uzyskać wynik dla jednej liczby. Oznacza to, że możliwości komputera kwantowego są dużo większe niż klasycznego. Pewne wymagania stawiane podstawowym elementom logicznym w komputerach kwantowych mogą być spełnione jedynie wtedy, gdy elementy te będą wykorzystywać zjawiska kwantowe [41], np. te zachodzące w nadprzewodnikach. Pierwsze próby budowy kwantowych bramek logicznych na bazie nadprzewodników zostały poczynione i wypadły pomyślnie.

## 9. Co dalej?

Obecnie znamy dużą liczbę (ponad 2000) związków i stopów wykazujących nadprzewodnictwo. Ich liczba wciąż rośnie. Za badania związane bezpośrednio z nadprzewodnictwem przyznano 4 Nagrody Nobla. Badania niskotemperaturowe, włączając w to zjawisko nadciekłości, uhonorowano ogółem siedmioma Nagrodami Nobla. Co dalej? Przewidywanie przyszłości nie jest rutynowym zajęciem fizyków. Można się jednak pokusić o pewne ogólne stwierdzenia. Nie ma najmniejszej wątpliwości, że trzy główne nurty badań występujące w fizyce, tzn. badania doświadczalne, próby ich zrozumienia i opisu teoretycznego oraz próby zastosowania uzyskanych wyników w praktyce będą w najbliższych kilkunastu latach bardzo intensywnie prowadzone.

Nie ma najmniejszej wątpliwości, że zostaną odkryte nowe nadprzewodzące materiały. Być może któryś z nich okaże się być nadprzewodnikiem w temperaturach pokojowych, tzn. w okolicach punktu zamarzania wody. Sądzę, że odkrywca takiego materiału ma szansę na Nagrodę Nobla. Odrębną kwestią będzie, czy taki materiał będzie dawał szanse na praktyczne zastosowania. Jak już mówiłem, wciąż nie rozumiemy właściwości nowych nadprzewodników. Fizycy nie są zgodni, jaki jest mechanizm zjawiska w nadprzewodnikach wysokotemperaturowych miedziowych i żelazowych. Pewne argumenty sugerują, że ten sam mechanizm jest odpowiedzialny za nadprzewodnictwo w materiałach organicznych i ciężko-fermionowych. Sformułowanie jednolitej teorii opisującej różne aspekty zjawiska w nowych, jak to się często mówi egzotycznych nadprzewodnikach jest, moim zdaniem osiągnięciem bardzo trudnym o ile w ogóle możliwym.

Analiza rozwoju fizyki ciała stałego i wynikających z niej praktycznych zastosowań wskazuje na jeden ważny kierunek badań. Są to badania układów mezoskopowych i nanoskopowych. Pojęcie "układ mezoskopowy" oznacza system fizyczny o rozmiarach pośrednich pomiędzy rozmiarami atomu a rozmiarami makroświata. Typowe rozmiary takich próbek są rzędu mikro- lub nawet nanometra, a więc jednej milionowej lub miliardowej metra. Ostatnio w wielu laboratoriach na świecie badane są miniaturowe nadprzewodniki o rozmiarach liniowych mikrometra lub mniejszych. W takich kropkach (lub punktach kwantowych) wykonanych z nadprzewodników należy oczekiwać zupełnie nowych i trudnych do przewidzenia zjawisk kwantowych. Czy ten kierunek poszukiwań zakończy się spektakularnymi zastosowaniami nadprzewodników? Obecnie tego jeszcze nie wiemy. To musi być sprawdzone.

## 10. Podsumowanie

Na zakończenie chciałbym wyraźnie zaznaczyć, że w tym podsumowaniu specjalnie nie poświęciłem wiele miejsca odkryciom ostatnich trzydziestu lat. Do tego okresu nie mam (nie mamy) właściwego dystansu. Uważam, że na takie podsumowanie przyjdzie jeszcze czas. Będzie ono jednak bardzo trudne, gdyż po odkryciu nadprzewodnictwa ciężko fermionowego, wysokotemperaturowego w związkach miedzi i żelaza, w materiałach organicznych, w kryształach bez środka symetrii czy nadprzewodnictwa trypletowego pojawiło się wiele nowych prób teoretycznego opisu wszystkich tych związków. I choć idea parowania fermionów i łamania symetrii cechowania U(1) wciąż pozostaje podstawowym elementem wszystkich propozycji teoretycznych, to inne elementy teorii są często zupełnie nowe. Co więcej, teorie te wciąż są na etapie udoskonalania i doświadczalnego sprawdzania ich przewidywań.

## Literatura